\documentclass[floats,twocolumn,showpacs,superscriptaddress]{revtex4-1}

\usepackage{graphicx}% Include figure files
\usepackage{rotate}
\usepackage{epsfig}
\usepackage{xcolor}
\usepackage[normalem]{ulem}
\usepackage{amssymb,amsfonts,amsmath}

\definecolor{MyBrick}{rgb}{0.84,0.01,0.01}

\newcommand{\avg}[1]{\langle #1 \rangle}

\newcommand{\etal}{\emph{et al.} }
\newcommand{\ie}{\emph{i.e.} }

\graphicspath{{./pictures/}}

\begin{document}

\title{Coevolution of synchronization and cooperation in costly networked interactions}
%\title{Coevolution of synchronization and cooperation in networks of coupled oscillators}

\author{Alberto Antonioni}
\email{alberto.antonioni@unil.ch}
\affiliation{Faculty of Business and Economics, University of Lausanne, CH-1015~Lausanne, Switzerland}
\affiliation{Grupo Interdisciplinar de Sistemas Complejos (GISC), Departamento de Matem\'aticas, Universidad Carlos III de Madrid, E-28911~Legan\'es, Madrid, Spain}
\affiliation{Institute for Biocomputation and Physics of Complex Systems (BIFI), University of Zaragoza, E-50018 Zaragoza, Spain}

\author{Alessio Cardillo}
\email{alessio.cardillo@epfl.ch}
\affiliation{Laboratory for Statistical Biophysics, \'Ecole Polytechnique F\'ed\'erale de Lausanne (EPFL), CH-1015 Lausanne, Switzerland}
\affiliation{Institute for Biocomputation and Physics of Complex Systems (BIFI), University of Zaragoza, E-50018 Zaragoza, Spain}


% POSSIBILI PACS CANDIDATES 
% 89.75.-k 	Complex systems 
% 05.45.Xt 	Synchronization; coupled oscillators
% 02.50.Le 	Decision theory and game theory
%
% OTHER CANDIDATES
%
% 87.23.Ge 	Dynamics of social systems
% 89.75.Fb 	Structures and organization in complex systems
% 05.70.Fh 	Phase transitions: general studies
% 

\begin{abstract}

Despite the large number of studies on synchronization, the hypothesis that interactions bear a cost for involved individuals has been considered seldom. The introduction of costly interactions leads, instead, to the formulation of a dichotomous scenario in which an individual may decide to cooperate and pay the cost in order to get synchronized with the rest of the population. Alternatively, the same individual can decide to \emph{free ride}, without incurring in any cost, waiting that others get synchronized to her state. The emergence of synchronization may thus be seen as the byproduct of an evolutionary game in which individuals decide their behavior according to the benefit/cost ratio they accrue in the past. We study the onset of cooperation/synchronization in networked populations of Kuramoto oscillators and report how topology is essential in order for cooperation to thrive. We display also how different classes of topology foster differently synchronization both at a microscopic and macroscopic level.

\end{abstract}

\maketitle

% INTRODUZIONE

The simultaneous occurrence of events known as \emph{synchronization} constitutes one of the most fascinating phenomenon ever studied. Synchronization has, in fact, been observed in systems of different nature from power grids, to biological and chemical environments just to name a few \cite{pikovsky-book-2003, motter-nphys-2013, varela-nat_rev_neur-2001, mirollo-siam-1990}. If the system is composed of many units, the pattern of interactions among them plays a cornerstone role in the establishment of the conditions under which the onset of synchronization occurs~\cite{arenas-physrep-2008,rodrigues-physrep-2015}. Initially, scientists have studied such onset by encoding interactions as either a mean-field or a regular lattice \cite{mirollo-siam-1990, pikovsky-book-2003}. However, interactions in real systems are better described as \emph{complex networks} \cite{boccaletti-physrep-2006} and, therefore, it is important to pursue the study of synchronization within such paradigm \cite{arenas-physrep-2008}.\\
\indent Despite the abundance of studies regarding the emergence of synchronization in networked populations of dynamical units, a key aspect has been neglected thus far: the existence of a \emph{cost} associated to the interactions responsible for the update of the state of each dynamical unit. Yet, it seems reasonable to assume that the introduction of a cost affects the dynamics. With the present Letter we sought to highlight what happens to the synchronization phenomena when interactions between individuals are regulated by an evolutionary non-cooperative game. More specifically, we are interested in studying the emergence of synchronization in systems where agents can decide whether to interact -- or not -- with their neighbors by evaluating the cost sustained to alter their own state and the benefit received by getting more synchronized with the rest of the neighborhood. Decision processes based on the evaluation of a payoff constitute the heart of evolutionary game theory \cite{maynard_smith-book-1982, blume-games-1993}. Thus, a \emph{coevolutionary} approach based on synchronization and evolutionary game theory is the natural frameset to study this kind of problem. We are interested in assessing under which conditions the will of getting synchronized (\ie cooperating) affects the attainment of a global synchronized state. In addition, since the topology of interactions plays a role of paramount importance for the onset of a collective behavior \cite{arenas-physrep-2008, barrat-book-2008, roca-physlrev-2009}, we also want to study the nexus between different topologies and the thriving of cooperation/synchronization.\\
\indent In the proposed model we study the emergence of synchronization and cooperation in networked populations of coupled Kuramoto oscillators \cite{kuramoto-1984, strogatz-physd-2000, acebron-rev_mod_phys-2005, rodrigues-physrep-2015}. Each node corresponds to a different agent/oscillator called to face a dilemma, named the \emph{Evolutionary Kuramoto's Dilemma} (EKD), whenever it must decide the strategy to adopt in the future. As we will show, the structure of interactions affects dramatically the conditions under which cooperation and synchronization thrive. Specifically, we observe the onset of a new transition from the synchronized to incoherent state when either the coupling strength or the relative cost become too elevate. Moreover, the location of the transition non-trivially depends on the topology of the interaction structure.

%\section{The model}

%\subsection{Synchronization}

Let us consider a graph $G(N,\avg{k})$ of $N$ nodes and average degree $\avg{k}$. Each node, $l$, represents a dynamical unit and its state is characterized by its \emph{strategy} $s_l$, \emph{phase} $\theta_l$, and \emph{natural frequency} $\omega_l$. The strategy is equal to $s_l=1$ ($s_l=0$) if the agent is a cooperator (defector) and shapes the interaction pattern of the agent. The evolution of the phase is governed by a slightly modified version of the Kuramoto model \cite{kuramoto-1984, strogatz-physd-2000, acebron-rev_mod_phys-2005} given by:
\begin{equation}
\label{eq:kuramoto-coop}
\dot{\theta_l} = \omega_l + s_l \, \lambda \sum_{j=1}^N a_{lj} \, \sin(\theta_j - \theta_l)\,,
\end{equation}
where $\lambda$ is the coupling strength, and $a_{lj}$ are the elements of the adjacency matrix $\mathcal{A}$ of $G$. Initially, both $\theta$ and $\omega$ are uniformly distributed over the interval $[-\pi,\pi]$. To gauge the global synchronization level of the system, we use the \emph{Kuramoto order parameter} $r_G$~\cite{kuramoto-1984, strogatz-physd-2000}, given by:
\begin{equation} 
\label{eq:orderparam}
r_G \, e^{i\Psi} = \dfrac{1}{N} \sum_{j=1}^N e^{i \theta_j} \qquad \text{ with } r_G \in [0,1]\,,
\end{equation}
where $i$ is the imaginary unit and $\Psi$ is the average phase of the system. When $r_G = 0$ the system is in the incoherent state where all the oscillators have distinct phases. Conversely, $r_G = 1$ denotes a fully coherent (\ie synchronized) state where all the oscillators have the same phase. Apart from the global order parameter, it is possible to define a \emph{local} measure, $r_l$, which only accounts for the level of synchronization of a given node with respect to its neighbors. This, in turns, stems from the general case of Eq.~\eqref{eq:orderparam}, which for a \emph{pair} of nodes $l$ and $m$ is:
\begin{equation}
\label{eq:pairwise-order}
r_{lm} \, e^{i(\theta_l+\theta_m)/2} =  \, \dfrac{e^{i\theta_l}+e^{i\theta_m}}{2}\,.
\end{equation}
Hence, for a node $l$, the local order parameter, $r_l$, is:
\begin{equation}
\label{eq:localorderparam} 
r_l = \dfrac{\sum_{m=1}^{N} a_{lm} \,  r_{lm}}{\sum_{m=1}^N a_{lm}} \,,
\end{equation}
Consequently, we can define its average over all the nodes as $r_L = \tfrac{1}{N} \sum_l r_l$.

%
%\subsection{Evolutionary game}
%

The evolution of the strategy of a given node/player $l$ depends on the evaluation of the \emph{payoff} accumulated during one discrete step of synchronization dynamics. The payoff is given by the difference between the benefit $b_l$ and the cost $c_l$, that the player attains. The former accounts how much an oscillator has converged towards being synchronized with its neighbors and is equal to the local order parameter $r_l$, given by Eq.~\eqref{eq:localorderparam}. The latter is given by the absolute value of the angular acceleration, \ie:
\begin{equation}
\label{eq:cost}
c_l = \Delta \dot{\theta_l} \equiv \left\lvert \dot{\theta_l}(t) - \dot{\theta_l}(t-\epsilon) \right\rvert \,,
\end{equation}
where $\epsilon=0.01$ is the discrete step used to compute the synchronization dynamics given by Eq.~(\ref{eq:kuramoto-coop}) with the Runge-Kutta method.
The absolute value is required to ensure that $c_l$ is always semidefinite positive. Finally, the payoff of a node is defined as:
\begin{equation}
\label{eq:payoff}
\Pi_l = r_l - \alpha \dfrac{c_l}{2\pi} \,. 
\end{equation}
Here the cost has been divided by $2\pi$ to make it commensurable with the benefit. Furthermore, we modulate the role of the cost by multiplying it for a scalar $\alpha$ named \emph{relative cost}. The payoff function is designed in a way such that the benefit complies with the assumption that nodes with larger degree could accumulate a higher payoff, and also that low levels of synchronization may lead to negative payoffs. Once the players accumulate their payoffs, they decide the strategy to adopt in the next time step by means of the so-called \emph{Fermi rule}~\cite{blume-games-1993, szabo-pre-1998, roca-physlrev-2009}. In such rule, the focal player $l$ randomly selects one of its neighbors, $m$, and adopts its strategy with a probability given by:
\begin{equation}
\label{eq:fermirule}
P({s_l \leftarrow s_m}) = \dfrac{1}{1 + e^{-\beta (\Pi_m - \Pi_l)}}\,,
\end{equation}
where $\beta$ accounts for the ``irrationality'' of the players. Without loss of generality, we use $\beta = 1$ throughout this study. We also point out that our results are qualitatively in agreement with those using other update rules~\cite{roca-physlrev-2009}, see Supplementary Information (SI), Sec.~\ref{sec:other_updates}. We consider networks with $N=1000$ and $\avg{k}=6$. As initial condition, we set half of the population made of cooperators. After each synchronization step, agents accumulate their payoffs and synchronously update their strategies according to Eq.~(\ref{eq:fermirule}). We repeat these steps until the system reaches the stationary state.

To summarize, a cooperator always accepts to interact with its neighbors trying to reach mutual synchronization by paying the cost associated with such interactions. A defector, instead, refuses to interact with its neighbors and does not pay any cost. Given this scenario, we can consider a well-mixed population initially composed by $(N-1)$ cooperators plus a single defector. From an evolutionary point of view, the defector is highly advantaged obtaining, on average, the same benefit of all other members (getting synchronized) without incurring any cost. Thus, the defector has a payoff always greater than a cooperator, despite cooperating would imply a higher benefit for the entire population. Hence, the system evolutionarily undergoes a transition to complete defection which in the jargon of evolutionary game theory is called ``\emph{the tragedy of the commons}'' \cite{maynard_smith-book-1982,hardin-science-1968}.

%
%  FIGURA FENOMENI GLOBALI
%
%
\begin{figure*}
\centering
\includegraphics[width=0.8\textwidth]{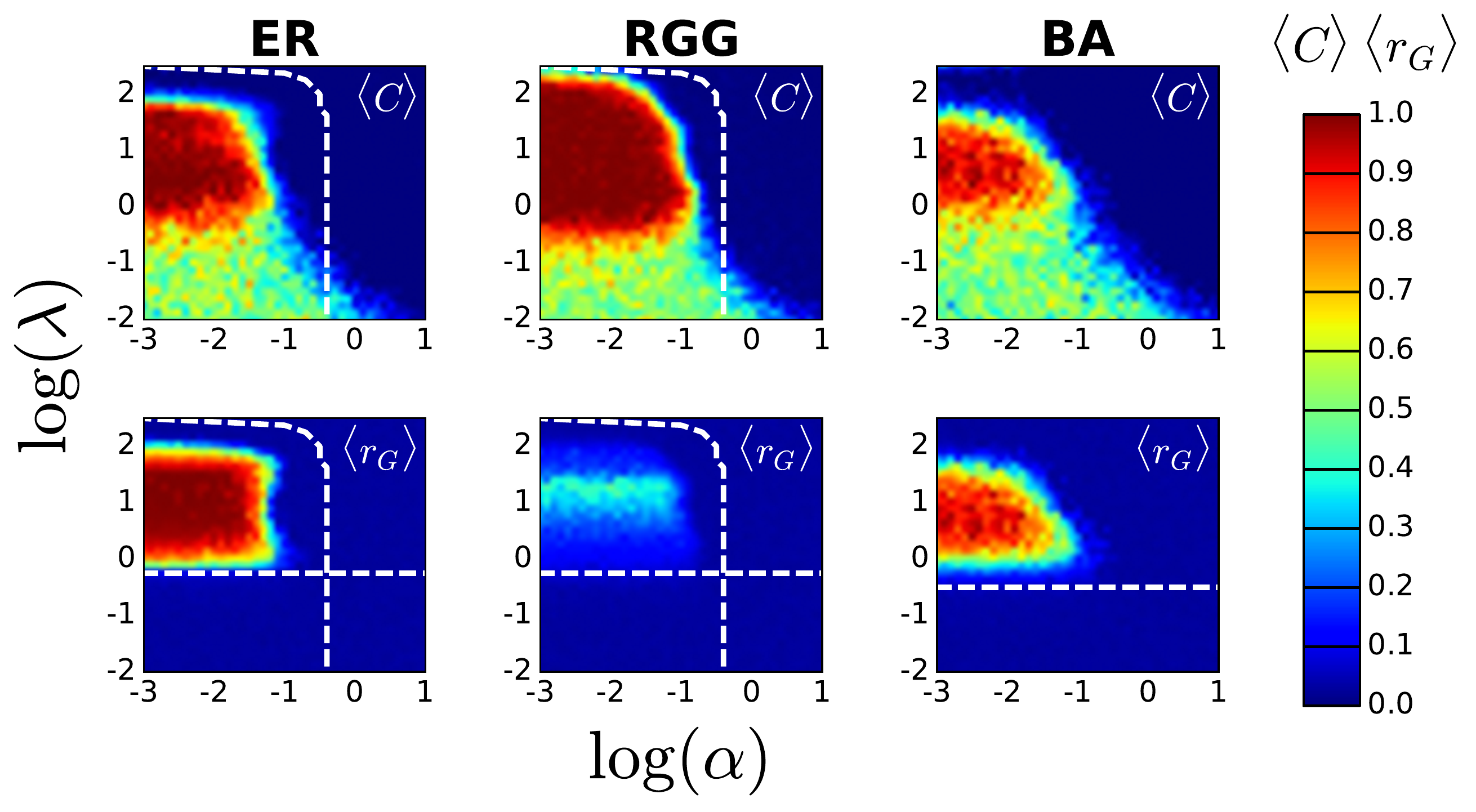}
\caption{(color online) Emergence of cooperation/synchronization at global scale. The top (bottom) row illustrates the average level of cooperation (synchronization) $\avg{C}$ ($\avg{r_G}$) as a function of the coupling $\lambda$ and relative cost $\alpha$. Each column corresponds to a different topology, namely: ER, RGG and BA. Results are averages over 50 different realizations.}
\label{fig:global}
\end{figure*}
%

%\section{Results}

As we have seen, the well-mixed scenario leads to the emergence of a fully defection and incoherent state. On the other hand, it has been reckoned that the structure of interactions is one of the mechanisms fostering the emergence of both cooperation and synchronization \cite{nowak-science-2006, gomez_gardenes-prl-2007}. In the light of that, we can sought if networked interactions may overcome the temptation to defect and, in turns, lead to the emergence of a coherent state. It is therefore interesting to study under which circumstances networked populations allow synchronization to thrive. To explore such scenario, we choose three different network topologies, namely; Erd\H{o}s-R\'enyi (ER) random graphs, Random Geometric Graphs (RGG) and Barab\'asi-Albert (BA) scale-free networks~\cite{erdos-1960, dall-pre-2002, barabasi-science-1999, boccaletti-physrep-2006}. 
%
%
%  FIGURA FENOMENI LOCALI
%
%
\begin{figure*}
\centering
\includegraphics[width=0.8\textwidth]{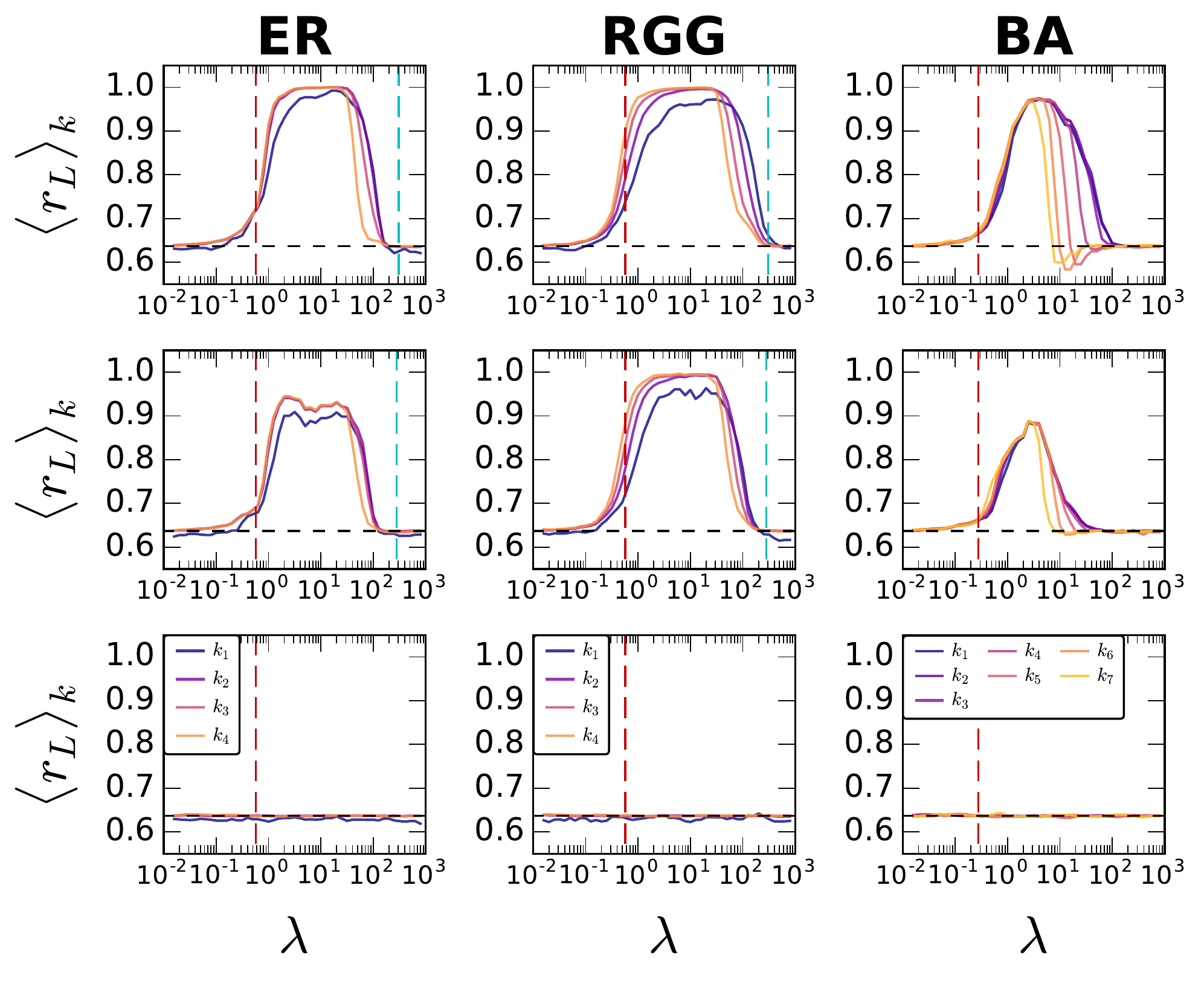}
\caption{(color online) Emergence of cooperation/synchronization at microscopic scale. Each plot displays the average local Kuramoto parameter for different degree classes $\avg{r_L}_{k_\sigma}$ as a function of the coupling $\lambda$ and for the three studied topologies (ER, RGG, BA). Colors refer to different degree classes: 
$k_\sigma \!\!\in\!\! \lbrace k_1,\ldots,k_4 \Longleftrightarrow [1,3], [4,7], [8,10], [11,\infty] \rbrace$ for both ER and RGG, and $k_\sigma \!\!\in\!\! \lbrace k_1,\ldots,k_7 \Longleftrightarrow [1,5], [6,10], [11,20], [21,40], [41,55], [56,80], [81,\infty] \rbrace$ for BA. Each row corresponds to a different value of $\alpha$:  top row, $\alpha = 10^{-3}$, middle row, $\alpha = 10^{-1.4}$, and bottom row, $\alpha = 1.0$.  Results are averages over 50 different realizations.}
\label{fig:local_rl}
\end{figure*}

We begin the study of our toy model by looking at its global behaviour. In Fig.~\ref{fig:global} we display both the average level of cooperation, $\avg{C}$ (top), and the global order parameter, $\avg{r_G}$ (bottom), as a function of the relative cost $\alpha$ and coupling strength $\lambda$. At first glance, we notice a strong correlation between the onsets of cooperation and synchronization due to the intertwining among these dynamics. For example, for low and intermediate values of $\alpha$, the system displays a transition from an incoherent state to a fully coherent one when the coupling exceeds a critical value. It is worth mentioning that such transition takes place at couplings slightly above the expected ones given by $\lambda_c^{\text{theo}} = \lambda_c^{\text{MF}} \tfrac{\avg{k}}{\avg{k^2}}$, where $\lambda_c^{\text{MF}}$ is the critical value of the coupling in the mean field case~\cite{gomez_gardenes-prl-2007}. For each topology, we display $\lambda_c^{\text{theo}}$ as an horizontal dashed line. For low relative costs, as the coupling increases, we observe the appearance of a second spontaneous transition from a coherent state to an incoherent one. The reason behind such loss of synchronization is the increasing burden associated with the variation of the angular speed, \ie the cost, with respect to the benefit associated to the increase of synchronization. Such transition is observed exclusively for small to intermediate values of $\alpha$ and the region where synchronization takes place shrinks as $\alpha$ increases. Ultimately, for high values of $\alpha$ synchronization never occurs independently on the coupling strength and on the considered topology. In the case of homogeneous networks, following the method introduced by Ohtsuki \etal \cite{ohtuski-nature-2006} (see SI, Sec.~\ref{sec:theory_ohtsuki}), we can analytically perimeter  the region of the parameter space inside which a single defector is able to invade a population of cooperators. Such region is delimited by the curved dashed line in the ER and RGG panels. In analogy with other studies on coevolutionary dynamics~\cite{leventhal-nat_comm-2015}, the results shown in Fig.~\ref{fig:global} tell us that BA networks do not promote synchronization and cooperation more than ER random graphs; in spite of what reported previously for each dynamics~\cite{gomez_gardenes-prl-2007,arenas-physrep-2008,santos2005, santos2006}. Yet, RGG panels point towards the existence of additional mechanisms associated with spatial correlations and the presence of community structure in order to converge to more cooperative outcomes (see SI, Sec.~\ref{sec:rgg}).

\indent To shed light on such discrepancy, in Fig.~\ref{fig:local_rl} we present a microscopic analysis by measuring $\avg{r_L}_{k_\sigma}$ for different connectivity classes, $k_\sigma$, and for three scenarios of relative cost. More precisely, we consider the following cases: $\alpha = 10^{-3}, 10^{-1.4}, 1.0$, to account for cheap, intermediate and expensive interactions, respectively. Darker solid lines account for low degree nodes while brighter lines account for highly connected ones. The vertical dashed lines delimit the region where synchronization theoretically emerges (dark red) and vanishes (cyan). The horizontal line, instead, accounts for the analytical value that $r_L$ assumes when two nodes are randomly selected in a well-mixed population (see SI, Sec.~\ref{sec:avg_rlm}) accounting for a scenario where no synchronization is observed. A closer inspection to Fig.~\ref{fig:local_rl} reveals a set of intriguing features. The first is that in BA networks -- and to large extent also in ER ones -- nodes tend to attain synchronization all together, in accordance with \cite{gomez_gardenes-prl-2007}, but then they lose it in a hierarchical order, instead. This is not the case for the RGG where a hierarchy exists in both transitions, whose quantitative analysis deserves further investigation. Moreover, there is a value of $\lambda$ for which the hierarchy gets inverted meaning that nodes placed at the core of communities tend to act as condensation nuclei exerting a positive feedback for the formation of cooperative clusters, but are also burdened with higher costs and therefore lose synchronization sooner (see SI, Sec.~\ref{sec:rgg}). The loss of synchronization is particularly strong in BA networks where is usually accompanied by a drop of $r_L$ below the average value in the cheap interactions regime. The intermediate cost regime presents a similar picture but where the second transition shrinks. Finally, in the high cost regime we observe the complete absence of synchronization also at a local level. The average cooperation level essentially follows the same trend of the local synchronization (see SI, Sec.~\ref{sec:micro_strategy}). We also performed simulations to test the robustness of our results for different initial conditions, letting the system synchronizing for $\tau$ steps before starting the coevolutionary process and a higher initial fraction of cooperators (see SI, Sec.~\ref{sec:other_initial}).\newline
%
%
%\section{Conclusions}
\indent In conclusion, by accounting for the existence of a cost associated to synchronization interactions, we have built a coevolutionary toy model based on the intertwining of synchronization and evolutionary game theory. The resulting dilemma has been named the \emph{Evolutionary Kuramoto's Dilemma} (EKD). Network reciprocity has proven itself as a valuable catalyzer of both the survival of cooperation and synchronization, and the effects of accounting for the cost of interaction in networked populations are twofolds. On one hand, high values of the coupling result in the desynchronization of the system, which has been previously observed only in \emph{higher order} models like the R\"ossler \cite{yanchuk-physd-2001}. The other result is the appearance of a hierarchy in the degree of nodes attaining/losing synchronization, which depends on the topology in a non-trivial manner. Our numerical results have been performed for relatively small system size, although analogous chimera states can emerge also on larger RGG and lattice topologies~\cite{abrams2004,pannagio2015}. Considering more realistic patterns, like spatial proximity and heterogeneous ones, sheds some light on such dependence and paves the way to further investigations. Finally, the model presented here -- apart from looking at synchronization under a different perspective -- can be used to tackle a wide range of problems, leveraging the interplay between coupling, cost and topology to drive the system towards a coherent or incoherent state. For example, the relation between cooperation and synchronization has been explored in the context of Social and Cognitive/Behavioral Sciences. In the former, previous studies pointed at assessing whether the execution of coordinated/synchronized actions may foster the onset of cooperation (military drills are an example of such situations) \cite{wiltermuth-psycho-2009, lakens-soc_cogn-2011, bruggeman-arxiv-2015}. In the latter, instead, studies aimed at understanding how the individual \emph{behavioural plasticity} -- \ie the will of individuals to disregard their individual preferences -- could be altered to improve the synchronization \cite{yokoyama-plos_compbio-2011, herbert_read-proc_roy_soc_b-2012, alderisio-bio_cyb-2016, alderisio-arxiv-2016, slowinsky-jroysocint-2016}. An even more appealing application is represented by the so-called ``social insects'' (ants, bees, fireflies, etc.) whose synchronized (or not) behaviour -- \ie the flashing of fireflies and \emph{Leptothorax acervorum} ants activity cycle to cite a few -- can be explained in terms of \emph{natural selection} \cite{bonabeau-tren_ecol_evol-1997, sumpter-phil_tran_roy_soc-2006, hamilton-jtheob-1964, buck-sci_am-1976, boi-proc_roy_soc_b-1999}. Lastly, a modified version of the Kuramoto model has been used also to study the attainment of consensus in opinion dynamics~\cite{pluchino-ijmpc-2005, pluchino-physa-2006, castellano-rev_mod_phys-2009}.

%  ACKNOWLEDGEMENTS

The authors thank M. Frasca, J. G\'omez-Garde\~nes and P. De Los Rios for helpful discussions and valuable comments.
AA acknowledges the financial support of SNSF under grant no. P2LAP1-161864.
AC acknowledges the financial support of SNSF through the project CRSII2\_147609.

%%%%%%%%%%%%%%%%%%%%%%%%%%%%%%%%%%%%%%%%%%%%%%%%%%%%%%%%%%%%%%%
%
%   INIZIO SUPPLEMENTARY MATERIAL
%
%%%%%%%%%%%%%%%%%%%%%%%%%%%%%%%%%%%%%%%%%%%%%%%%%%%%%%%%%%%%%%%

\pagebreak
\widetext
\begin{center}
\textbf{\large Supplemental Information: \\Coevolution of synchronization and cooperation in costly networked interactions}
\end{center}
%%%%%%%%%% Merge with supplemental materials %%%%%%%%%%
%%%%%%%%%% Prefix a "S" to all equations, figures, tables and reset the counter %%%%%%%%%%
\setcounter{equation}{0}
\setcounter{figure}{0}
\setcounter{table}{0}
\setcounter{page}{1}
\setcounter{section}{0}
\makeatletter
\renewcommand{\theequation}{S\arabic{equation}}
\renewcommand{\thefigure}{S\arabic{figure}}
\renewcommand{\thesection}{\Roman{section}}   
%%%%%%%%%% Prefix a "S" to all equations, figures, tables and reset the counter %%%%%%%%%%

% \graphicspath{{./figures/}}

% include main text to correctly display references
% \externaldocument[M-]{main}% <- full or relative path 

%\hypersetup{
%pdfauthor={9210},
%pdftitle={},
%colorlinks,
%linktocpage=true, pdfstartpage=1, pdfstartview=FitV,
%breaklinks=true, pdfpagemode=UseNone, pageanchor=true, pdfpagemode=UseOutlines,
%plainpages=false, bookmarksnumbered, bookmarksopen=true, bookmarksopenlevel=1,
%hypertexnames=true, pdfhighlight=/O,
%urlcolor=RoyalBlue, linkcolor=RoyalBlue, citecolor=RoyalBlue}

% \newcommand{\HRule}{\rule{\linewidth}{0.5mm}}
% \renewcommand{\Pr}{\mathbb{P}}
% %\newcommand{\parder}[2]{ \frac{\partial #1}{\partial #2} }
% \newcommand{\comments}[1]{}
% 
% \renewcommand{\thetable}{S\arabic{table}}
% \renewcommand{\thefigure}{S\arabic{figure}}
% \renewcommand{\tablename}{Table}
% \renewcommand{\figurename}{Figure}
% 
% \definecolor{MyBrick}{rgb}{0.84,0.01,0.01}
% 
% \newcommand{\alessio}[1]{\textcolor{MyBrick}{#1}}
% 
% 
% \newcommand{\etal}{\emph{et al.} }
% \newcommand{\ie}{\emph{i.e.} }
% 
% 
% \clearpage
% \setcounter{page}{1}
% 
% % aumento della spaziatura tra le righe delle formule
% \setlength{\jot}{10pt}
% 
% 
% \makeindex
% 
% %%%%%%%%%%%%%%%%%%%%%%%%%%%%%%%%%%%%%%%%%%%%
% \begin{document}
% 
% \maketitle

% \section*{mean $r_L$}
\section{Average pairwise order parameter}
\label{sec:avg_rlm}

Considering a population of $N$ Kuramoto's phase oscillators, being $\theta_l$ and $\theta_m$ the phases of two randomly selected elements $l$ and $m$. Without loss of generality, we can rotate the reference system by an angle $-\theta_l$ such that $\theta^\prime_l = 0$ and $\theta^\prime_m = \theta_m - \theta_l = \theta$. Since $\theta$ is uniformly distributed over $[-\pi,\pi]$; we can compute the average value of the pairwise order parameter, $\overline{r_{lm}}$, between $l$ and $m$, given by Eq.~\eqref{eq:pairwise-order} in the main text as:
\begin{equation}
\label{eq:avg_rlm}
\begin{split}
\overline{r_{lm}} & = \; \dfrac{1}{2\pi} \int_{-\pi}^{\pi} \, \dfrac{\left\lVert 1+e^{i\theta} \right\rVert}{2} \, d\theta = \dfrac{1}{2\pi} \int_{-\pi}^{\pi} \, \dfrac{\left\lVert 1+\cos \theta+i\sin\theta \right\rVert}{2} \, d\theta= \\
& = \; \dfrac{1}{2\pi} \int_{-\pi}^{\pi} \, \dfrac{\sqrt{[1+\cos \theta]^2+\sin^2 \theta}}{2} \, d\theta = \dfrac{4}{2\pi} = \dfrac{2}{\pi} \sim 0.6366 \;.
\end{split}
\end{equation}

\section{Estimation of the conditions under which cooperation can emerge}
\label{sec:theory_ohtsuki}

The standard procedure to determine the conditions under which cooperation may thrive is to compute the difference of payoffs between a cooperator and a defector. If such difference is positive, the cooperator will be evolutionary advantaged with respect to the defector. Remembering that the payoff of a node $l$ is given by:
\begin{equation}
\label{eq:payoff}
\Pi_l = r_l - \alpha \dfrac{c_l}{2\pi} \,. 
\end{equation}
Let us consider a system of two oscillators $l$ and $m$ having $\theta_l$ and $\theta_m$ as phase and $\dot\theta_l$ and $\dot\theta_m$ as frequency, respectively. We want to calculate the benefit $b_C$ ($b_D$), \ie $r_{lm}$, when node $l$ is a cooperator (defector) and node $m$ is a defector. This means that node $m$ does not change its frequency $\dot\theta_m$, while node $l$ adjusts $\theta_l$ when behaves as a cooperator interacting with $m$. As commented in Sec.~\ref{sec:avg_rlm}, we consider $\theta_m=0$. For simplicity, we put $x= \theta_l$. When both oscillators are defectors, we have: 
\begin{equation}
\label{eq:ben_defectors}
\begin{split}
b_D &= \; r_{lm} = \dfrac{\left\lVert 1+e^{ix} \right\rVert}{2} = \dfrac{\left\lVert 1+\cos(x)+i\sin(x) \right\rVert}{2} = \\
 & = \;  \dfrac{\sqrt{[1+\cos(x)]^2+\sin^2(x)}}{2} = \dfrac{\sqrt{2[1+\cos(x)]}}{2} = \sqrt{\dfrac{1+\cos(x)}{2}} \,.
 \end{split}
\end{equation}

According to the Kuramoto's model, if $l$ is a cooperator it will change its phase from $\theta_l(t)$ to $\theta_l(t+\varepsilon)$. To highlight the role of the interactions in the process of accumulation of the payoff, we impose that the reference system turns at the same speed of $l$. Therefore, we can set $\omega_l = 0$ and write:
$$\theta_l(t+\varepsilon) = \theta_l(t) + \varepsilon\lambda\sin(\theta_m(t)-\theta_l(t)) \,.$$
where $\varepsilon$ represents the discrete time step size and $\lambda$ the coupling strength. Since $\theta_m = 0$, we obtain:
$$\theta_l(t+\varepsilon) = \theta_l(t) + \varepsilon\lambda\sin(-\theta_l(t)) =  x - \varepsilon\lambda\sin(x) \,.$$
Consequently:
\begin{equation}
\label{eq:ben_cooperators}
b_C = \; r_{lm} = \dfrac{\left\lVert 1+e^{i[x-\varepsilon\lambda\sin(x)]}\right\rVert}{2} = \sqrt{\dfrac{1+\cos(x-\varepsilon\lambda\sin(x))}{2}}\,.
\end{equation}

The difference in benefit $\Delta b=b_C-b_D$ gives the benefit increase $b$ that node $m$ receives from node $l$ when the latter cooperates. The cost, $c$, in which node $l$ incurs when cooperating is: \\

$c=\dfrac{\alpha\left\lvert\dot\theta_l(t+\varepsilon)-\dot\theta_l(t)\right\rvert}{2\pi} = \dfrac{\alpha \left\lvert \omega_l + \varepsilon\lambda\sin(x) - \omega_l \right\rvert}{2\pi}= \dfrac{\alpha \left\lvert \varepsilon\lambda\sin(x) \right\rvert}{2\pi}$.
\newline

As for Sec.~\ref{sec:avg_rlm}, we consider a mean-field scenario assuming that $\theta_m=0$ and $\theta_l= \tfrac{\pi}{2}$. Thus, we have:
\begin{align*}
b_C &= \sqrt{ \cfrac{ \biggl[ 1 + \cos \Bigl( \cfrac{\pi}{2} - \varepsilon \lambda \sin{ \cfrac{\pi}{2}} \Bigr)  \biggr]}{2}} =  \sqrt{ \cfrac{ 1 + \cos\Bigl( \cfrac{\pi}{2} - \varepsilon\lambda \Bigr)}{2}} = \sqrt{ \cfrac{ 1 + \sin\left( \varepsilon\lambda\right)}{2}} \,,\\
b_D & =\sqrt{ \dfrac{1 + \cos \dfrac{\pi}{2}}{2}} = \dfrac{\sqrt{2}}{2} \,.\\
\intertext{Therefore: }%\alessio{forse possiamo togliere il $b$ e mettere direttamente $\Delta b$}}
b &= \Delta b =  b_C-b_D=\sqrt{\dfrac{1+\sin(\varepsilon\lambda)}{2}}-\dfrac{\sqrt{2}}{2}= \dfrac{\sqrt{2+2\sin(\varepsilon\lambda)}-\sqrt{2}}{2}\,, \\
c &= \dfrac{\alpha \left\lvert \varepsilon\lambda\sin\dfrac{\pi}{2} \right\rvert}{2\pi}=\dfrac{\alpha \left\lvert \varepsilon\lambda \right\rvert}{2\pi}=\dfrac{\alpha\varepsilon\lambda}{2\pi}.
\end{align*}
\newline
Applying the method introduced by Ohtsuki \etal \cite{ohtuski-nature-2006} in the case of a homogeneous network with $N$ nodes and average degree $k$, we find that cooperators have a fixation probability greater than $1/N$, and defectors have a fixation probability less than $1/N$, if:
\begin{gather*}
b/c > k\\
\dfrac{\sqrt{2+2\sin(\varepsilon\lambda)}-\sqrt{2}}{\alpha\varepsilon\lambda}\pi > k\\
\dfrac{\sqrt{2+2\sin(\varepsilon\lambda)}-\sqrt{2}}{\varepsilon\lambda k}\pi > \alpha \,. 
\end{gather*}

The last inequality represents the mean-field condition in which cooperation can evolve and it is the curved line plotted in Fig.~\ref{fig:global} of the main text. We did not plot these condition for heterogeneous networks, as the BA scale free topology, since they could not be considered accurate enough.

\newpage
\section{Microscopic analysis of the strategy}
\label{sec:micro_strategy}

\begin{figure}[h]
\centering
\includegraphics[width=\textwidth]{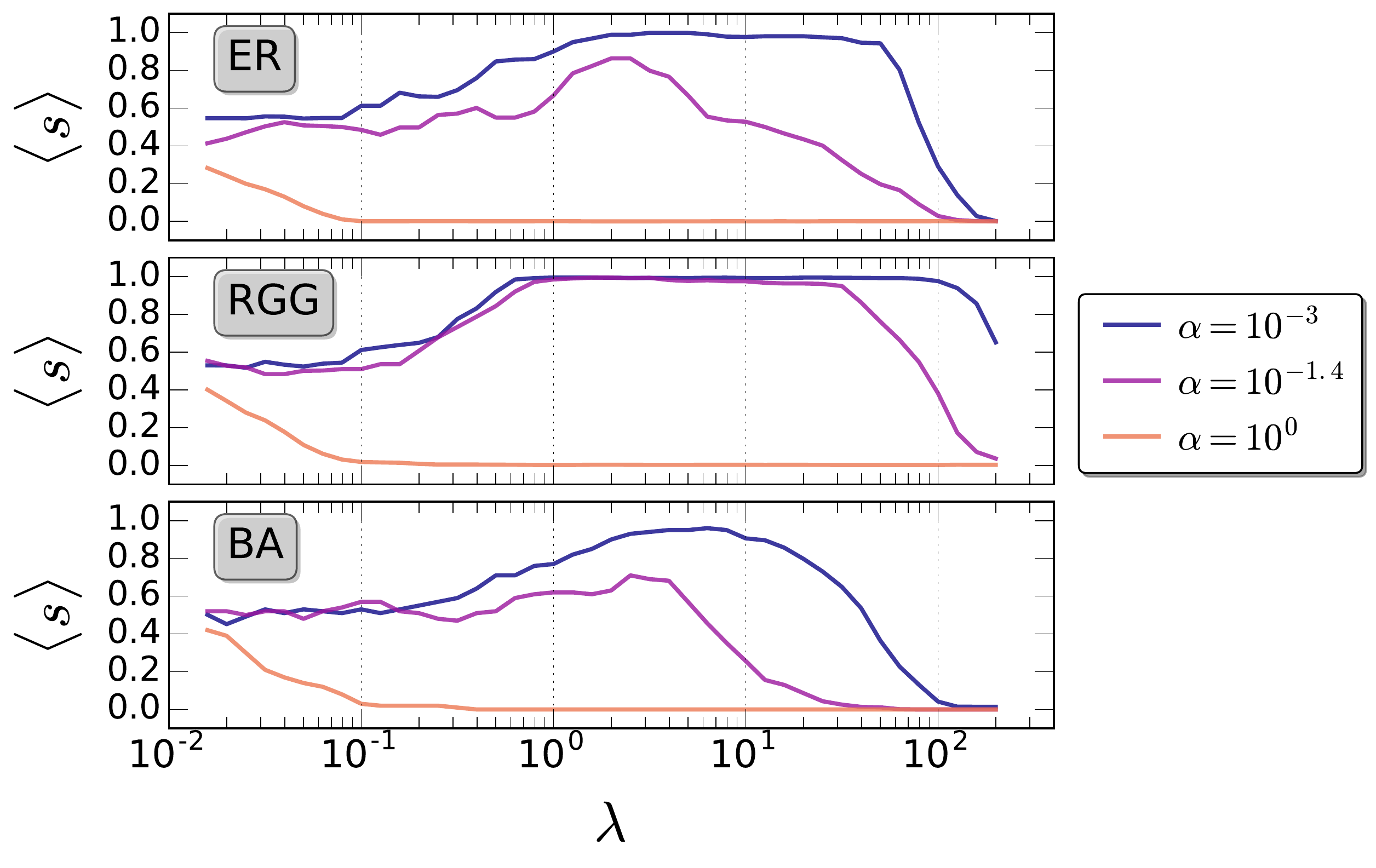}
\caption{Average strategy $\langle s \rangle$ as a function of the coupling $\lambda$ for all the considered topologies. For each topology, we display three values of relative cost $\alpha$, namely: $10^{-3}$, $10^{-1.4}$ and $10^{0}$. Results are averaged over 50 different realizations.}
\label{fig:micro_strategy}
\end{figure}

In Fig.~\ref{fig:micro_strategy} we display the strategy of the nodes in the stationary state as a function of the coupling $\lambda$ for the three cost regimes of the main manuscript. In contrast with Fig.~\ref{fig:local_rl}, we do not observe any difference associated to the degree. Such absence can be explained by the fact that these results are drawn in the stationary state where the population is generally formed by agents with the same strategy. As we will comment in the next section, there could be cases in which the stationary state may correspond to a population not fully made by agents with the same strategy. However, since the results displayed in Fig.~\ref{fig:micro_strategy} are averaged over 50 different realizations, those ``anomalies'' are smoothed away by averaging over the ensemble of realizations. \newline

\newpage
\section{Evolution of cooperation in Random Geometric Graphs}
\label{sec:rgg}

\begin{figure}[h!]
\centering
\includegraphics[width=\textwidth]{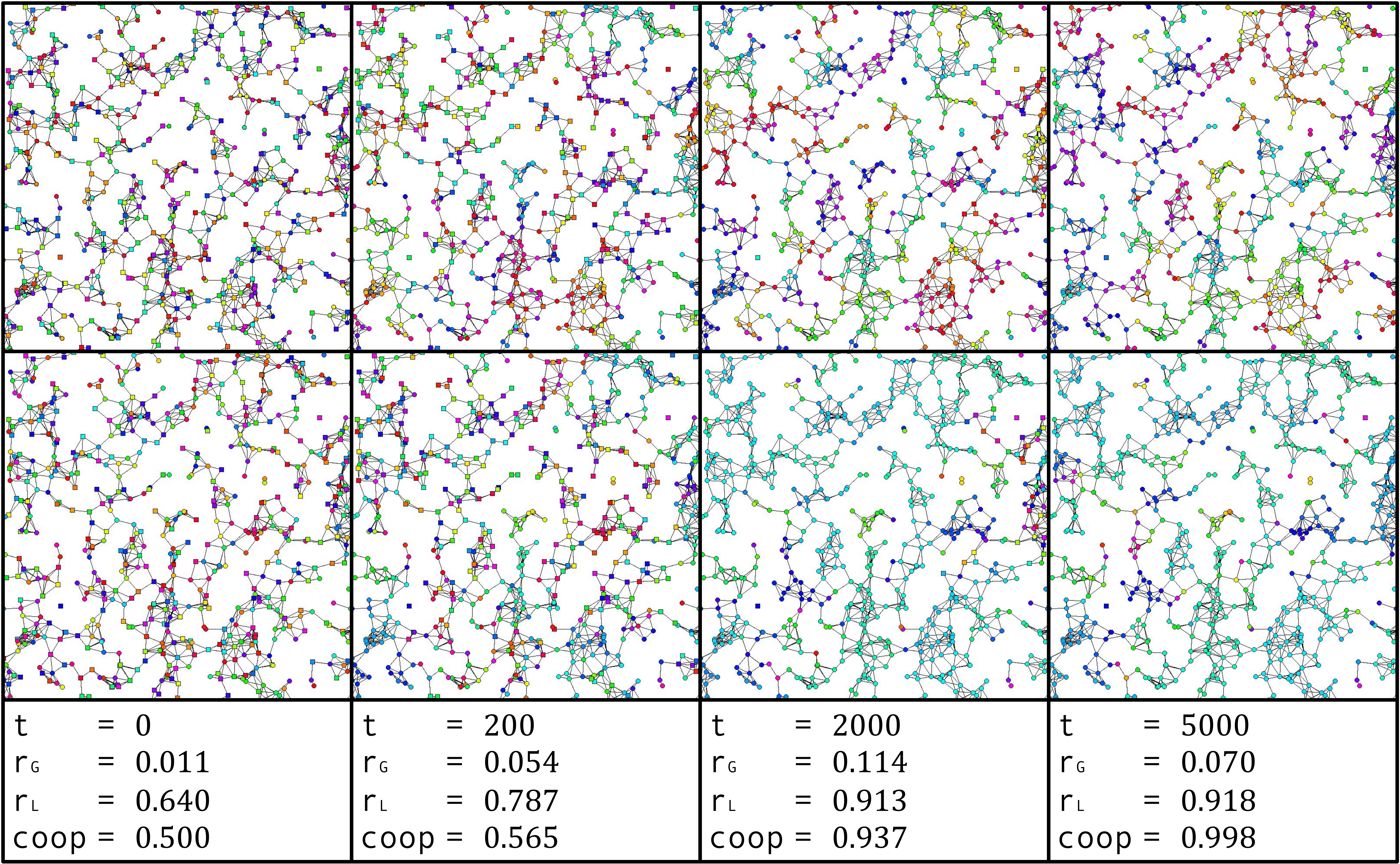}
\caption{Evolution in time and space of cooperation and synchronization in the cheap regime ($\lambda=10^0, \alpha=10^{-3}$). The nodes are represented as circles (squares) if they are cooperators (defectors). The color encodes the phase $\theta$ (top row) and angular speed $\dot{\theta}$ (middle) using the HSB scale. From left to right, each panel corresponds to time $t$ equals to 0, 200, 2000 and 5000 steps respectively. The nodes are placed on a torus (boundaries connections are not displayed).}
\label{fig:rgg_cheap}
\end{figure}

\begin{figure}[h!]
\centering
\includegraphics[width=\textwidth]{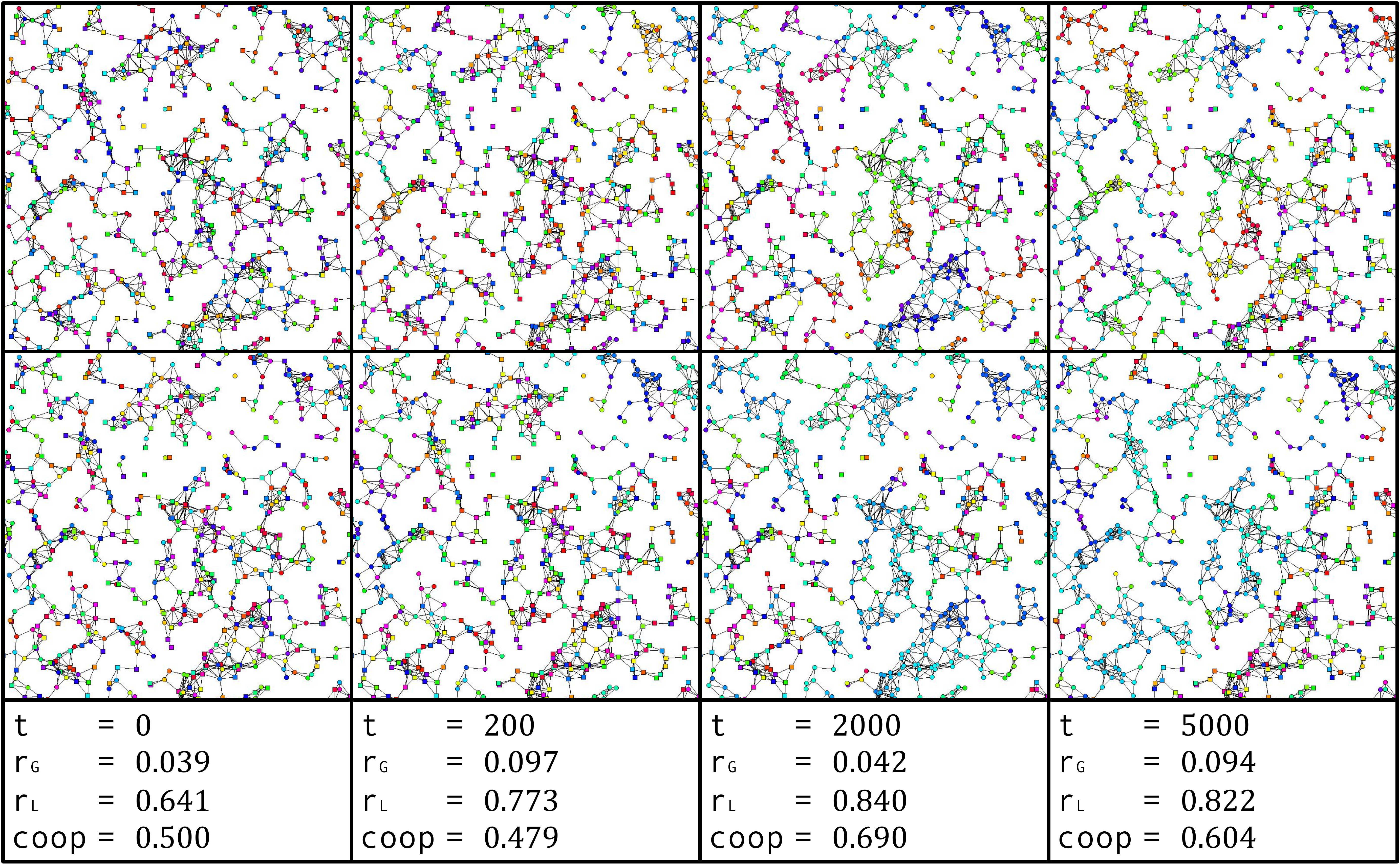}
\caption{Evolution in time and space of cooperation and synchronization in the intermediate regime: $\lambda=10^0$, $\alpha=10^{-0.8}$. All the other settings are the same as Fig.~\ref{fig:rgg_cheap}.}
\label{fig:rgg_intermed}
\end{figure}

\begin{figure}[h!]
\centering
\includegraphics[width=\textwidth]{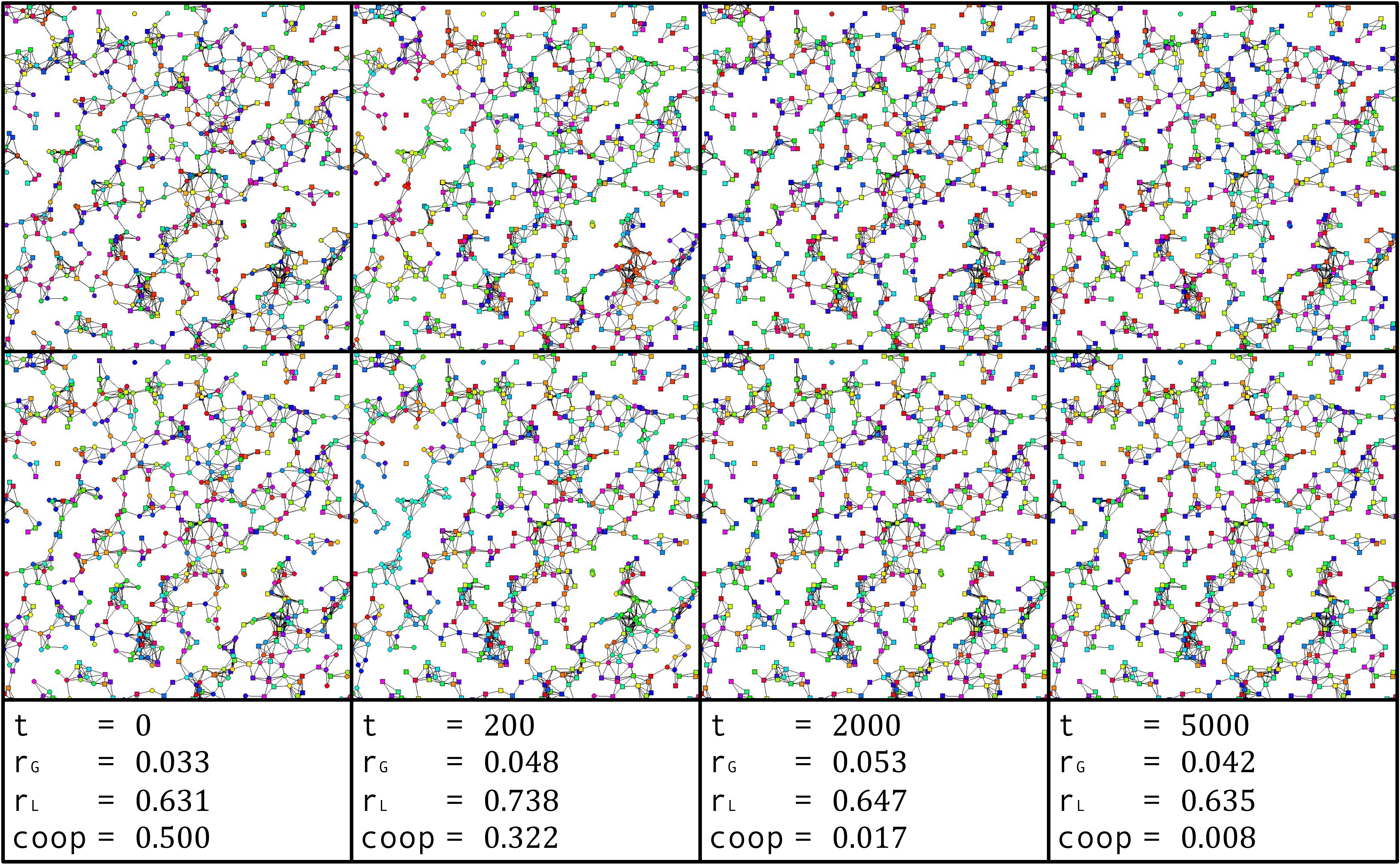}
\caption{Evolution in time and space of cooperation and synchronization in the expensive regime: $\lambda=10^0$, $\alpha=10^{-0.5}$. All the other settings are the same as Fig.~\ref{fig:rgg_cheap}.}
\label{fig:rgg_expensive}
\end{figure}

We want to display the evolution of cooperation in the case of a RGG topology. In Figs.~\ref{fig:rgg_cheap} -- \ref{fig:rgg_expensive} we display in color code the phase $\theta$, and angular speed $\dot{\theta}$ for all the $N$ nodes in the system at four different instants in time $t$. Each figure refers to $\lambda=10^0$ and a different relative cost regime, namely: \emph{cheap} (\ref{fig:rgg_cheap}, $\alpha=10^{-3}$), \emph{intermediate} (\ref{fig:rgg_intermed}, $\alpha=10^{-0.8}$) and \emph{expensive} (\ref{fig:rgg_expensive}, $\alpha=10^{-0.5}$), respectively. In the same pictures we display also the global and local levels of synchronization together with the fraction of cooperators in the population. 

In the cheap regime, Fig.~\ref{fig:rgg_cheap}, from $t=2000$ onwards, the system is almost fully made of cooperators. Given so, one may conclude that the global synchronization level $r_G$ should also converge to one, but, this is not the case. The spatial structure of the interactions bolsters the formation of small highly synchronized groups -- also confirmed by the value of $r_L$ -- that do not get synchronized with each other. Thus, the system ends in a phase-locked state corroborated by the fact that $\dot{\theta}\sim0$ for all oscillators (middle panel of Fig.~\ref{fig:rgg_cheap}). This value falls exactly at the middle of the range $[-\pi, \pi]$ which is the range used to initialize both $\theta$ and $\dot{\theta}$. 

The intermediate regime, Fig.~\ref{fig:rgg_intermed}, shows the emergence, almost since the very beginning, of a cluster of defectors -- represented as squared nodes -- able to resist the invasion of cooperators, and causing the decrease of the synchronization level around its periphery. Such destructive effect is enough to prejudicate completely the global synchronization which, in fact, is well approximated by an incoherent state ($r_G \simeq 0$).
\newpage

The expensive cost regime picture, Fig.~\ref{fig:rgg_expensive}, portraits the utter defeat of cooperation. The interactions are, indeed, so expensive that cooperators connected with defectors immediately change their strategy, paving the way to the invasion of defectors. In this scenario, cooperation dies out almost immediately (not shown) blasting away any chance of attaining synchronization.

\newpage

\newpage
\section{Results for other initial conditions}
\label{sec:other_initial}

\begin{figure}[h!]
\centering
\includegraphics[width=\textwidth]{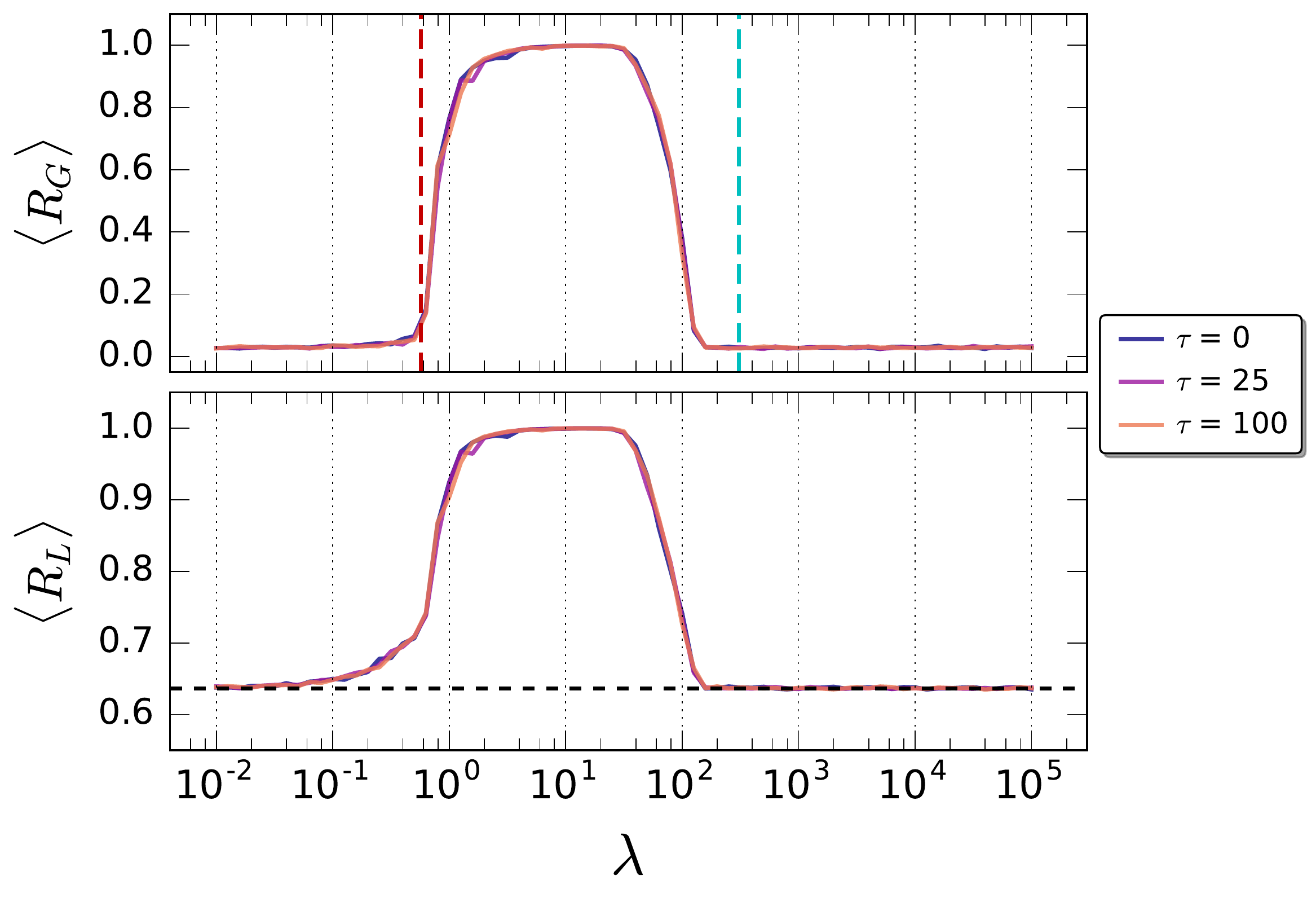}
\caption{Global $\langle R_G \rangle$ (top) and local $\langle R_L \rangle$ (bottom) synchronization level as a function of the coupling $\lambda$ for different times of first strategy update $\tau$. All results consider a relative cost $\alpha = 0.001$ and ER topology. Results are averaged over 50 different realizations.}
\label{fig:tvariation}
\end{figure}

To assess the robustness of the transition from the coherent to incoherent state, we alter the time at which nodes update their strategies for the first time. More specifically, the system evolves for a number of steps $\tau$ after which agents are allowed to update their strategy. The introduction of a transient time allows to prove if the emergence of a completely defective state is due to the fact that nodes are initially poorly synchronized and therefore accumulate little payoffs paving the way for the invasion of defectors. By letting the nodes synchronize for an amount of time $\tau$, we reduce such possibility. In Fig.~\ref{fig:tvariation} we plot the global $\langle R_G \rangle$ and local $\langle R_L \rangle$ order parameters as a function of the coupling $\lambda$ in the cheapest regime ($\alpha = 0.001$) for ER networks. As we can notice, changing the value of the first iteration at which update of strategies takes place does not produce any qualitative modification in the overall behavior of the system compared to the case where the update of the strategy takes place from the very beginning.

\begin{figure}[h!]
\centering
\includegraphics[width=\textwidth]{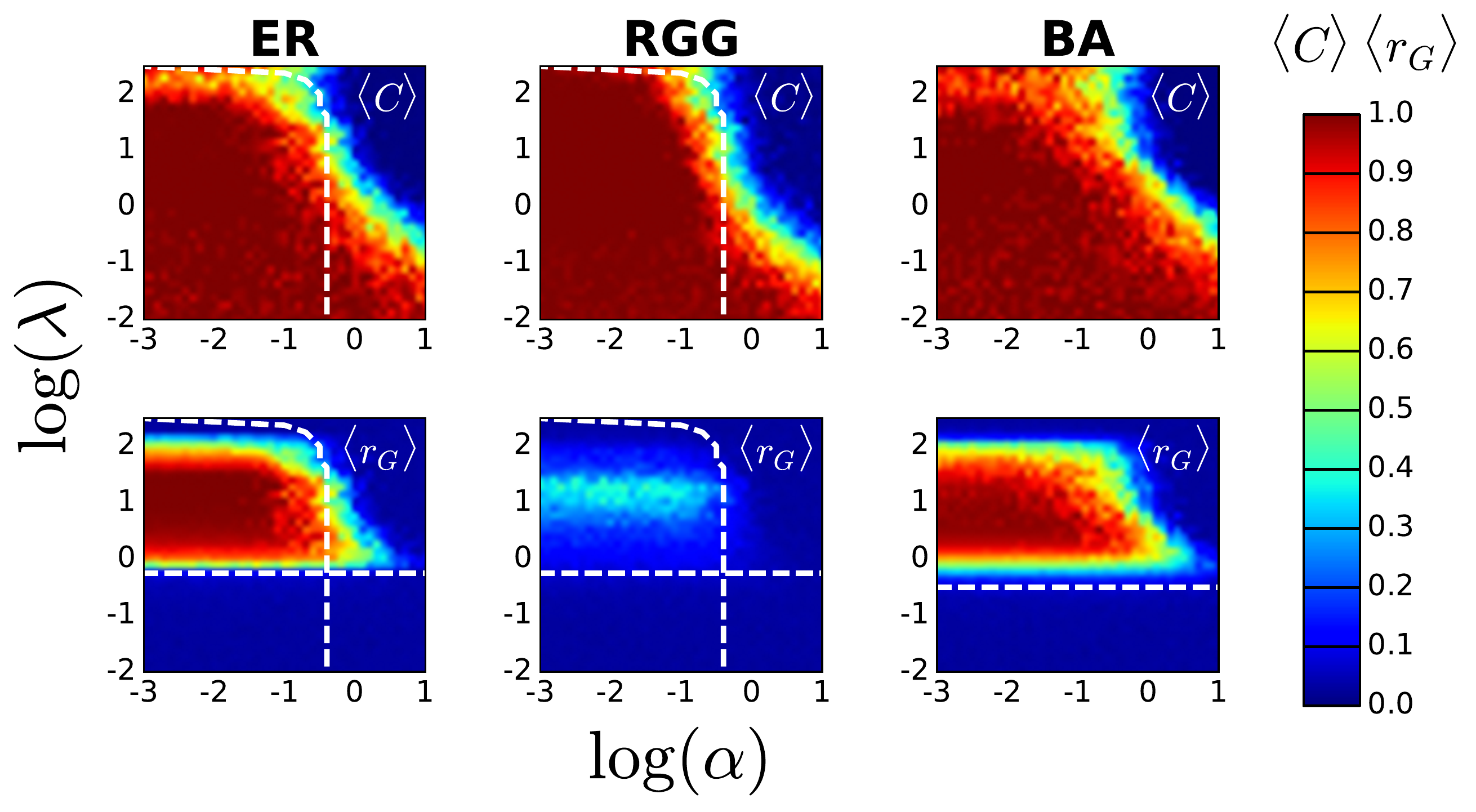}
\caption{Emergence of cooperation/synchronization at global scale for an initial population with $99\%$ of cooperators. The top (bottom) row illustrates the average level of cooperation (synchronization) $\avg{C}$ ($\avg{r_G}$) as a function of the coupling $\lambda$ and relative cost $\alpha$. Each column corresponds to a different topology, namely: ER, RGG and BA. Results are averages over 50 different realizations.}
\label{fig:coop-99}
\end{figure}

Figure~\ref{fig:coop-99} shows the same analysis of Fig.~\ref{fig:global} for a different initial fraction of cooperators ($99\%$). We observe an overall higher level of cooperation, also in regions in which cooperation was not previously observed and filling entirely the analytical region. As a byproduct, synchronization gets also enhanced but albeit we have cooperation also for couplings smaller than the critical one we cannot reach a synchronized state.

\newpage

\section{Other update rules}
\label{sec:other_updates}

\begin{figure}[h!]
\centering
\includegraphics[width=\textwidth]{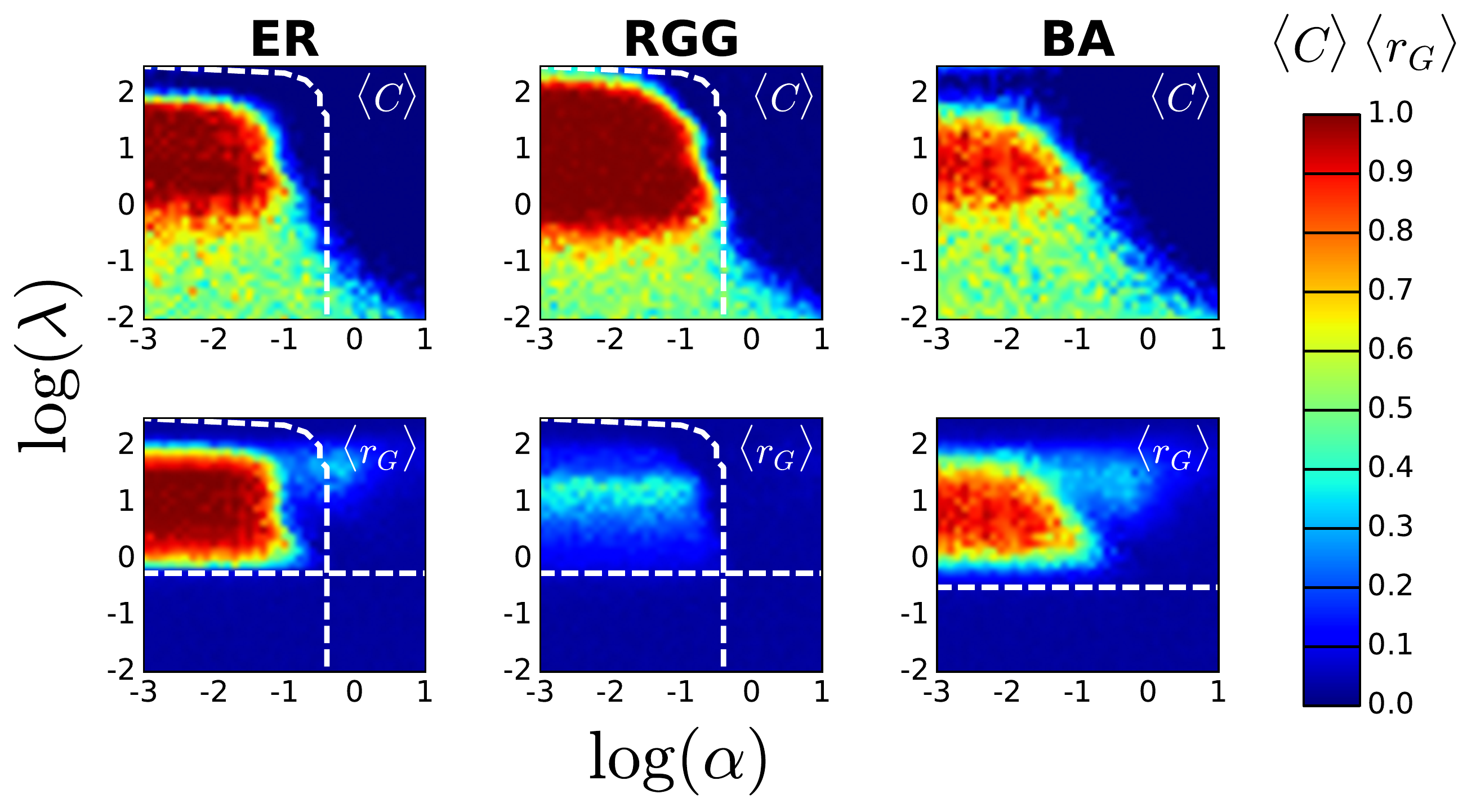}
\caption{Emergence of cooperation/synchronization at global scale for the asynchronous Fermi update rule. The top (bottom) row illustrates the average level of cooperation (synchronization) $\avg{C}$ ($\avg{r_G}$) as a function of the coupling $\lambda$ and relative cost $\alpha$. Each column corresponds to a different topology, namely: ER, RGG and BA. Results are averages over 50 different realizations.}
\label{fig:fermi-asynch}
\end{figure}

\begin{figure}[h!]
\centering
\includegraphics[width=\textwidth]{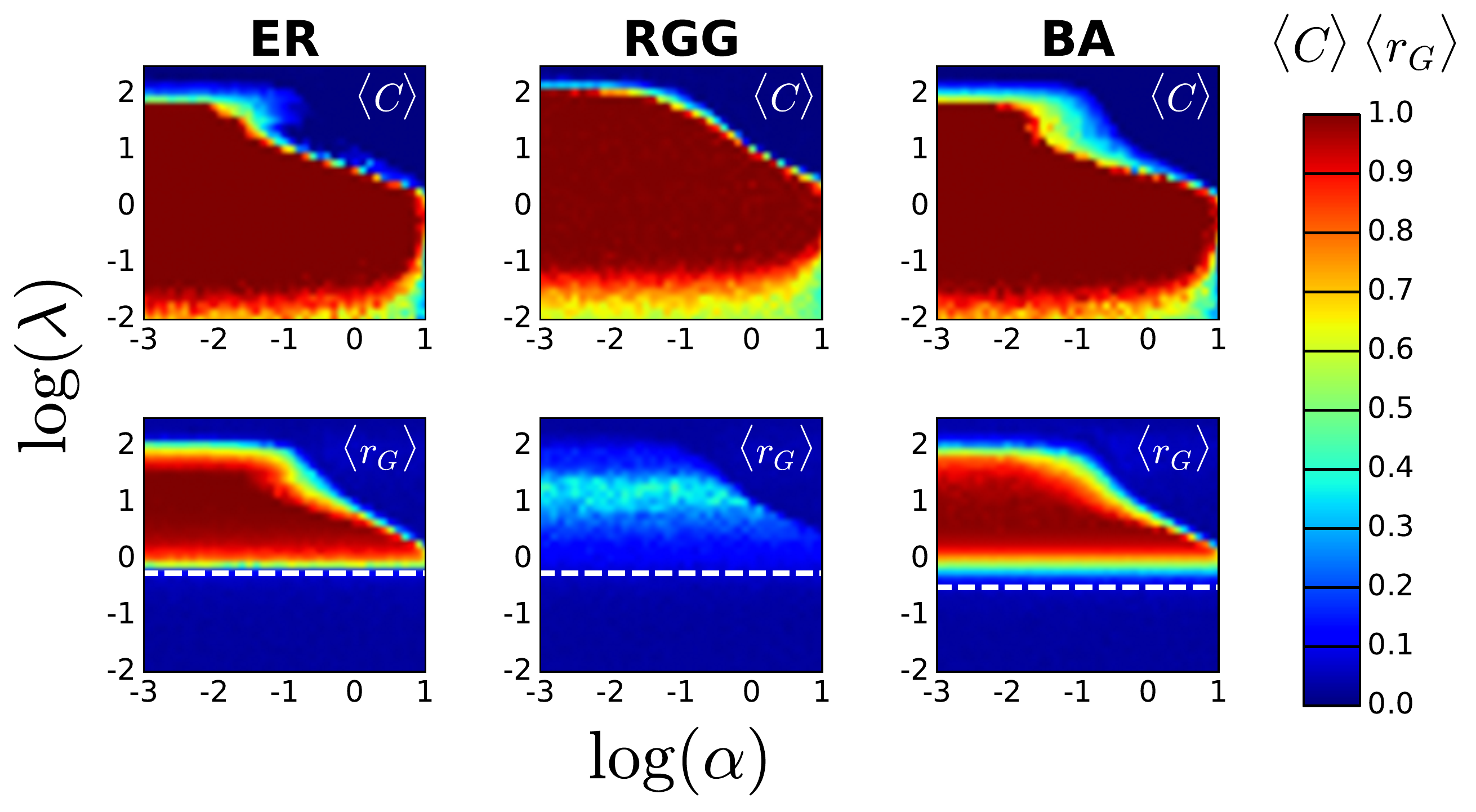}
\caption{Emergence of cooperation/synchronization at global scale for the synchronous Unconditional Imitation update rule. All the other settings are the same as in Fig.~\ref{fig:fermi-asynch}.}
\label{fig:imi-synch}
\end{figure}

In this section we study the effects of using update strategies other than the synchronous Fermi rule (Eq.~\eqref{eq:fermirule}) adopted throughout the main text. In Figs.~\ref{fig:fermi-asynch} and \ref{fig:imi-synch}, we present the same type of plots of Fig.~\ref{fig:global} but for the case of the asynchronous Fermi (AF) rule first, and synchronous Unconditional Imitation (UI) then.

In many biological contexts, the assumption that decisions are made in synchrony is unrealistic. Therefore, we relax such hypothesis allowing agents to change their behavior in a asynchronous way, i.e. at each time step only a single randomly chosen agent updates her strategy. Nevertheless, the overall levels of cooperation, $\avg{C}$, and global synchronization, $\avg{r_G}$, attained in all three network topologies remain pretty the same. We are thus able to conclude that the order of strategy updates seems to not play a significative role in the emergence of cooperation/synchronization.

In Fig.~\ref{fig:imi-synch} we explore the possibility to use a deterministic update rule. According to the UI rule, in fact, an agent $i$ updates her strategy $s_i$ by copying the one of her neighbor, $j$, having the highest payoff among all the neighbors of $i$, $N_i$ if and only if such payoff is higher than her own. Thus:
\begin{equation}
\label{eq:imitation}
s_i (t+1) = %
\left\{
\begin{aligned}
s_i (t)& &\text{ if } \Pi_i \geq \max\limits_{l \in N_i} \Pi_l\,,\\
s_j (t)& \text{ with } j \;\vert\; \Pi_j = \max\limits_{l \in N_i} \Pi_l \, &\text{ otherwise}.
\end{aligned}
\right.
\end{equation}

Regardless of the topology, the levels of cooperation and synchronization attained with UI are higher than in the Fermi rule. The reason of such enhancement is due to the propensity of UI to promote more cooperation as reported in \cite{roca-physlrev-2009}. However, despite the broader area where synchronization occurs, the qualitative behavior of the dynamics remains the same and we still observe the transition from coherence to incoherence as we increase either the coupling $\lambda$, the cost $\alpha$ or both. 
%

%
% \begin{thebibliography}{99}
% %
% \bibitem{ohtuski-nature-2006}
% Ohtsuki, H., Hauert, C., Lieberman, E.,  \& Nowak, M.~A. A Simple Rule for the Evolution of Cooperation on Graphs
% {\em Nature},  {\bf 441} 502--505, (2006).
% %
% % review on games on networks
% 
% \bibitem{roca-physlrev-2009} 
% Roca, C. P., Cuesta, J. a, \& S\'anchez, A. (2009). Evolutionary game theory: Temporal and spatial effects beyond replicator dynamics. Phys. Life Rev., {\bf 6}(4), 208--249, (2009).
% 
% \end{thebibliography}
% %%

\end{document}